\def\breakon{\end{multicols}\widetext\vspace{-.6cm}
\noindent\rule{.49\linewidth}{.3mm}\rule{.3mm}{.5cm}\vspace{0.0cm}}
\def\breakoff{\vspace{-0.45cm}
\noindent
\rule{.50\linewidth}{.0mm}\rule[-.47cm]{.3mm}{.5cm}\rule{.49\linewidth}{.3mm}
\vspace{-0.25cm}
\begin{multicols}{2}   }
\documentstyle[aps,epsf,prb,multicol]{revtex}
\newcommand{\be}{\begin{equation}}
\newcommand{\ee}{\end{equation}}
\newcommand{\bea}{\begin{eqnarray}}
\newcommand{\eea}{\end{eqnarray}}

\begin{document}

\makeatletter
\renewenvironment{table}
  {\let\@capwidth\linewidth\def\@captype{table}}
  {}

\renewenvironment{figure}
  {\let\@capwidth\linewidth\def\@captype{figure}}
  {}
\makeatother

\title{Phase Diagram of the spin $S=1/2$ Extended $XY$ model}
\author{Irakli Titvinidze and G.I. Japaridze}
\address{Andronikashvili Institute of Physics, Georgian Academy of Sciences\\
Tamarashvili str. 6, 380077, Tbilisi, Georgia}          
\maketitle

\begin{abstract}

The quantum phase transition in the ground state of the extended spin $S=1/2$ $XY$ 
model has been studied in detail.  Using the exact solution of the model the low temperature thermodynamics, 
as well as the ground state phase diagram of the model in the presence of applied uniform and/or 
staggered magnetic field are discussed.  
\end{abstract}


\vspace{.15in}
\begin{multicols}{2}

\section{\bf Introduction}

There is a considerable interest in models of strongly correlated electron and spin systems 
showing a Quantum Phase Transition (QPT) (see the Ref. \cite{Sachdev} ) . In the case of One-Dimensional 
electron or spin systems QPT related to the dynamical generation of a charge or spin gap 
is often connected with the change in the topology of the Fermi surface,
in particular with the doubling of the number of Fermi points \,\cite{Fabrizio96,Fabrizio98,DaulNoak00,Aebisher01}.

The one-dimensional spin $S=1/2$ $XY$ chain 
\be
{\cal H}_{XY}  = -J\sum_{n}\left(S^{x}_{n}S^{x}_{n+1} + S^{y}_{n}S^{y}_{n+1}\right) 
\ee
is the simplest exactly solvable strongly correlated spin model \,\cite{LSM,Katsura}.  
Its exact solution is expressed in terms of the {\em Fermi gas of spinless particles} 
(spinless Fermions (SF)). The free Fermi gas principle for construction of the system 
eigenstates and eigenvalues provides the straightforward and easy way to obtain the exact expressions for 
correlation functions and thermodynamic quantities  (see the Ref. \cite{Takahashi1} 
and references therein).

\begin{figure}
\epsfxsize=85mm
\centerline{\epsfbox{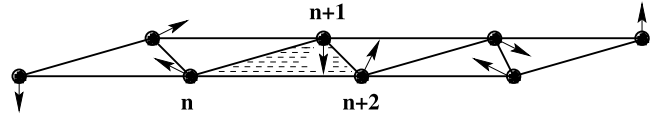} }
\vspace{5mm}
\caption{Schematic representation of the structure of the extended $XY$ model.}
\label{fig:Zig-zagladder}
\end{figure}

More than thirty years ago M. Suzuki proposed the whole class of generalized  $XY$ models with 
multi-spin interaction, allowing the exact solution in terms of the {\em Fermi gas of spinless Fermions} 
 \,\cite{Suzuki}. In the fermionic representation, the multi-spin coupling shows itself only through the form of the 
single particle spectrum. In this paper we consider the simplest of the proposed generalized $XY$ models - the 
extended $XY$ model with three spin coupling. The  Hamiltonian of the model is given by
\bea \label{GXYhamiltonian}
&{\cal H}  =   -J\sum_{n}\left(S^{x}_{n}S^{x}_{n+1} + 
S^{y}_{n}S^{y}_{n+1}\right) & \nonumber\\
 &  - J^{\ast}   \left(S^{x}_{n}S^{x}_{n+2} +
S^{y}_{n}S^{y}_{n+2}\right)S^{z}_{n+1} \,  &
\eea
and describes the spin system determined on the zig-zag chain (see Fig.\ref{fig:Zig-zagladder}),
with the transverse exchange between the spins on the nearest-neighbor sites $J$, and
transverse exchange between the spins on the next-nearest-neighbor sites $J^{\ast}$,
the latter depends on  ``z'' orientation
of the spin being between the  next-nearest-neighbors.
\vspace{5mm}
\vspace{3mm}

Despite its rather formal and possibly even nonrealistic nature the model  (\ref{GXYhamiltonian}) 
attracted our attention owing to its several important advantages:
\begin{itemize}
\item{it is an exactly solvable model;}
\item{the exact solution is expressed in terms of Fermi gas of SF.}
\end{itemize}
Moreover, as it is shown in this paper, the model (\ref{GXYhamiltonian}) is characterized by the rich ground state phase diagram. In particular
\begin{itemize} 
\item{with the increase of three spin coupling, at $J^{\ast}_{c} = 2J$ the system experiences QPT from the 
{\em Spin Liquid I} phase into the {\em Spin Liquid II} phase;}
\item{in the SF representation this QPT is associated with the doubling of Fermi points;}
\item{at the transition point the magnetic and low-temperature thermodynamic properties of the 
system show a well pronounced anomalous behavior;}
\item{in the case of applied uniform and/or staggered magnetic field the system is characterized by the 
rich ground state phase diagram  which contains {\em ferromagnetic, antiferromagnetic and three 
different spin liquid phases}.}
\end{itemize}

The paper is organized as follows. In the forthcoming Section we discuss the 
spinless Fermion representation of the model. In the Section III the ground state properties are
discussed. In the Section IV the low-temperature thermodynamics of the 
system is considered. In the Section V the ground state phase diagram
of the model is studied in the presence of magnetic field. Finally, the
Section VI contains the discussion and concluding remarks. 
The paper contains also two appendices. In Appendix A we present the table with 
exact expressions for the so called emptiness formation probability (EFP) $P(n)$ for 
$n=1,...,20$ and for 
different values of the parameter $J^{\ast}$. In Appendix B we present the expression for the 
EFP, which fits our exact data for $\alpha \geq \alpha_{c}$.

\section{Spinless Fermion representation}

The Hamiltonian (\ref{GXYhamiltonian}) can be diagonalized by means of the Jordan-Wigner 
transformation \,\cite{JW}
\begin{eqnarray} \label{JWtransformation}
\nonumber\\ 
S^{+}_{n}= S^{x}_{n} + {\it i}S^{y}_{n}&=&
\prod_{m=1}^{n-1}\left(1-2c_{m}^{\dagger}c_{m}\right)c_{n}^{\dagger}\nonumber\\
S^{-}_{n}= S^{x}_{n} - {\it i}S^{y}_{n} &=& \prod_{m=1}^{n-1}c_{n}
\left(1-2c_{m}^{\dagger}c_{m}\right)\\ 
S^{z}_{n} &=& c_{n}^{\dagger}c_{n} -\frac{1}{2} \nonumber 
\end{eqnarray}
where $c_{n}^{\dagger}, c_{n}$ are the spinless fermion (SF) creation and annihilation 
operators, respectively. In fact, the Hamiltonian (\ref{GXYhamiltonian}) is transformed 
to a free spinless fermion model as
\bea \label{GXYSFhamiltonian}
{\cal H} & = & -\frac{J}{2}\sum_{n}\left(c_{n}^{\dagger}c_{n+1}+ c^{\dagger}_{n+1}c_{n}\right)
\nonumber\\ 
&  +&  \frac{ J^{\ast}}{4}\sum_{n}\left(c_{n}^{\dagger}c_{n+2}+ c^{\dagger}_{n+2}c_{n}\right)\, . 
\eea

We can diagonalize the Hamiltonian (\ref{GXYSFhamiltonian}) by means of the Fourier 
transformation to obtain 
\begin{equation} \label{GXYSFhamiltonianEk}
{\cal H}=\sum_{k}E(k)c^{\dagger}(k)c(k)
\end{equation} 
where
\be \label{Spectrum}
E(k)=-J\left(\cos(k)-\frac{\alpha}{2}\cos2k\right).
\ee
and $\alpha=J^{\ast}/J$. 

In what follows we assume that $J>0$ and $\alpha>0$. However, since the unitary transformation 
\be \label{transformation1}
S_{n}^{x,y} \rightarrow (-1)^{n}S_{n}^{x,y}\, ,   \qquad  S_{n}^{z} \rightarrow S_{n}^{z} 
\ee
changes the sign of the nearest neighbor exchange constant $J \rightarrow - J$, one can easily 
reconstruct the features of the system for $J<0$ using the transformation (\ref{transformation1}). 
Moreover, since the sign of the three spin coupling term is changed by the time reversal transformation
\be \label{transformation2}
S_{n}^{y} \rightarrow S_{n}^{y}\, ,  \qquad  S_{n}^{x,z} \rightarrow -S_{n}^{x,z}\, ,   
\ee
the properties of the system for $\alpha< 0$ can be easily obtained from the ground state phase 
diagram of the system at $\alpha>0$ using the transformation (\ref{transformation2}).
\begin{figure}
\epsfxsize=85mm
\centerline{\epsfbox{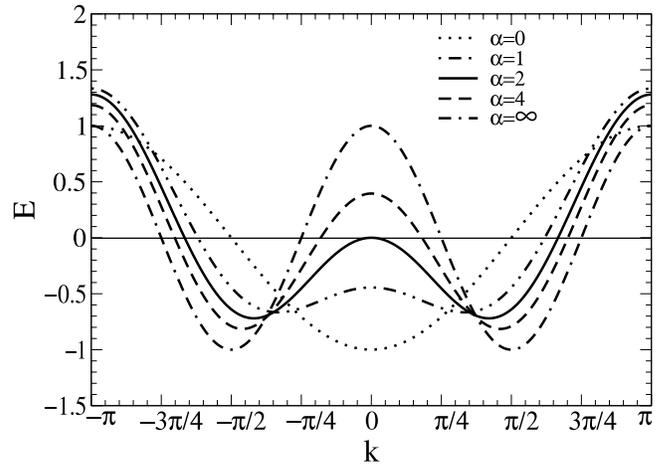} }
\vspace{5mm}
\caption{The spectra for different values of the parameter
$\alpha$. For given $\alpha$ the corresponding curve is scaled so as  
to give the same bandwidth equal to 2.}
\label{fig:Spectra}
\end{figure}
\vspace{3mm}

In the thermodynamic limit ($N \rightarrow \infty$), the ground state 
of the system corresponds to the configuration where all the states with $E(k) \leq 0$
are filled and $E(k)>0$ are empty. For $\alpha < 0.5$, the band minimum is at $k=0$ 
(see Fig.\ref{fig:Spectra}). If $\alpha > 0.5$, the band minima 
$\pm k_{\em min}$ move away from $k=0$ and satisfy the relation 
$\cos(k_{\em min}) = 1/2\alpha$, while a band maximum at $k=0$ is   
$E(0)=J\left(\alpha/2 - 1\right)$.  Therefore, for $\alpha < 2$ there
are only two Fermi points at 
\begin{equation} \label{kF1}
\pm k^{-}_{F}=  \arccos\left[\left(1-\sqrt{1+2\alpha^{2}}\right)/2\alpha\right] 
\end{equation} 
and the ground state corresponds to the configuration when all states
with $|k|<k^{-}_{F}$ are filled. At $\alpha = \alpha_{c}=2$ the band
maximum at $k=0$ reaches the Fermi level (see Fig. \ref{fig:Spectra})
and at $\alpha > \alpha_{c}$ $ E(0)>0$ and therefore, two additional 
Fermi points appear at  
\begin{equation}\label{kF2}
\pm k^{+}_{F}= \arccos\left[\left(1+\sqrt{1+2\alpha^{2}}\right)/2\alpha\right].
\end{equation}

In this case the ground state corresponds to the configuration when all states
with $k^{+}_{F} < |k| < k^{-}_{F}$ are filled.\\

As we show in this paper, the change of the topology of the Fermi surface 
of the equivalent SF model for 
$\alpha> \alpha_{c}=2$ corresponds to the phase transition in the ground state
of the spin system. This transition is the second order quantum phase transition, which at the 
transition point is characterized by 
\begin{itemize}
\item{non-monotonic behavior of the ground state energy $E_{0}(\alpha) $ as a function of the 
parameter  $\alpha$ . In particular, the first derivative of the ground state energy with 
respect to the parameter $\alpha$ shows a kink, while the second derivative is divergent 
at $\alpha = \alpha_{c}$; }
\item{magnetization of the system as a function of the parameter $\alpha$ shows the 
non-monotonic and nonanalytical behavior at $\alpha=\alpha_{c}$; }
\item{the critical index characterizing the power-law decay of the transverse spin-spin 
correlation function is changed at $\alpha=\alpha_{c}$;  }
\item{At low temperature ($T \ll J$) the heat capacity, magnetization and magnetic susceptibility of the system show a well pronounced anomalous 
behavior at $\alpha \simeq \alpha_{c}$;}
\end{itemize}

This quantum phase transition, despite its obviously rich nature, can be 
studied using the exact solution of the model (\ref{GXYhamiltonian}),
given in terms of {\em Fermi gas of spinless particles }  (\ref{GXYSFhamiltonianEk}).

\section{Phase transition in the ground state}

The ground state energy of the system is given by
\begin{equation} \label{GSE}
{\cal E}_{0}(\alpha)=\frac{L}{2\pi}\int_{\{\Lambda\}}{\cal E}(k)dk
\end{equation}
where the integration region $\{\Lambda\}=[-k^{-}_{F},k^{-}_{F}]$ for 
$\alpha <2$ and  
$\{\Lambda\}=[-k^{-}_{F},-k^{+}_{F}]\cup [k^{+}_{F},k^{-}_{F}]$  for 
$\alpha > 2$.

\begin{figure}
\epsfxsize=85mm
\centerline{\epsfbox{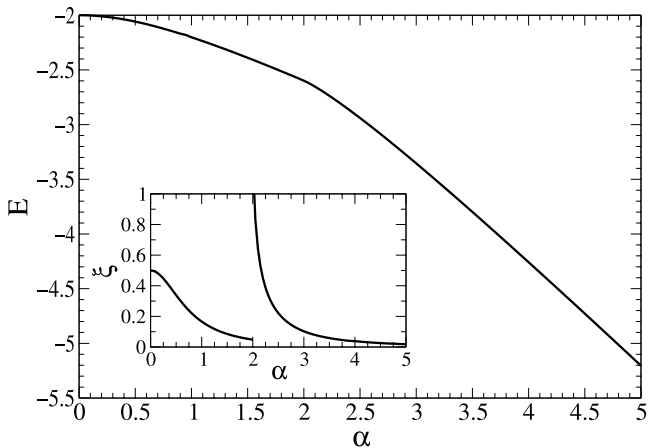} }
\vspace{0mm}
\caption{The ground state energy as a function of the parameter $\alpha$. The inset shows the  
generalized stiffness $\xi$ as a function of the parameter $\alpha$}
\label{fig:Energy}
\end{figure}

In Fig.\ref{fig:Energy} we plotted the ground-state energy of
the system as a function of the parameter $\alpha$. At $\alpha=\alpha_{c}$ we observe 
(see inset in Fig.\ref{fig:Energy}) the singularity in the behavior of the 
generalized stiffness 
$$
\xi(\alpha)=-\partial^{2}{\cal E}_{0}(\alpha)/\partial^{2}\alpha .
$$ 
The singularity in $\xi(\alpha)$ we attribute to the second-order
phase transition in the ground-state of the system with the increase of
the three-spin coupling constant $J^{\ast}=\alpha J$. 
\vspace{3mm}
\vspace{3mm}

\subsection{Magnetization}

Most clearly, the change of the properties of the
system at the transition point is seen in the non-monotonic behavior of magnetization.  
At $T=0$ the magnetization (per site) of the
system is given by the number of SF in the ground state
\bea\label{Magnetization}
&m^{z}(\alpha)=\frac{1}{L}\sum_{n}\langle S_{n}^{z}\rangle = &\nonumber\\
&=\left\{
\begin{array}{l}
k^{-}_{F}/\pi-1/2 \hskip1.7cm {\mbox at} 
\hskip0.2cm (\alpha < 2) \\
\left(k^{-}_{F}-k^{+}_{F}\right)/\pi - 1/2
\hskip0.3cm {\mbox at} \hskip0.2cm(\alpha > 2)
\end{array} 
\right. \,.  &
\eea 

\begin{figure}
\epsfxsize=85mm
\centerline{\epsfbox{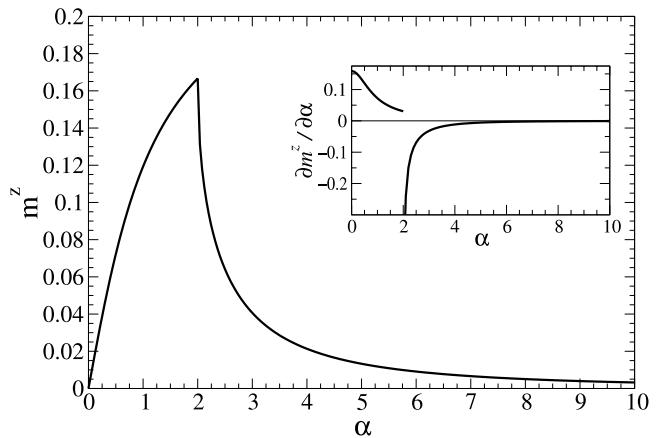} }
\vspace{5mm}
\caption{The ground state magnetization as a function of the parameter
$\alpha$. The inset shows the derivative of magnetization with respect to the 
parameter $\alpha$. } 
\label{fig:MegnetizationT_0}
\end{figure}

Using Eqs (\ref{kF1})-(\ref{kF2}) we found that the ground-state of the system is 
singlet with $m^{z}=0$ only at $\alpha=0$ (the XY model) and in the limiting case 
$\alpha=\infty$. For arbitrary nonzero $\alpha$, the magnetization of the system is finite. 
It is a monotonically increasing function of the parameter $\alpha$ in the range of 
$0 < \alpha < 2$ (see Fig.\ref{fig:MegnetizationT_0}). It reaches its cusp-type 
maximum at $\alpha=2$ with $m^{z}=1/6$ and then 
monotonically decreases to zero with $\alpha \rightarrow \infty$.

The derivative of magnetization with respect to the parameter $\alpha$ 
(see Inset in the Fig.\ref{fig:MegnetizationT_0}) is 
\bea
\partial m^{z}(\alpha)/\partial \alpha &\nonumber
\simeq
\left\{
\begin{array}{l}
{\em constant} \hskip0.8cm {\mbox at} 
\hskip0.3cm (\alpha-\alpha_{c}\rightarrow 0_{-}) \\
-\frac{1}{\sqrt{\alpha-\alpha_{c}}} \hskip0.8cm {\mbox at}
\hskip0.3cm  (\alpha-\alpha_{c}\rightarrow 0_{+})
\end{array} 
\right. \,. 
\eea

\vspace{3mm}

\subsection{Correlation functions}

To clarify the symmetry properties of various phases realized in
the ground state of the system, we calculate the longitudinal and 
transverse spin-spin correlation functions 
\bea
&{\cal K}^{z}(r)  =  \langle S_{n}^{z}S_{n+r}^{z} \rangle\, ,&\\
&{\cal K}^{\em tr}(r) = \langle S_{n}^{x}S_{n+r}^{x}\rangle = \langle S_{n}^{y}S_{n+r}^{y}\rangle\, .  &  
\eea

The spin-spin correlations in the case of ordinary $XY$ chain 
($\alpha=0$) were studied in the classical paper by Lieb, Schulz and
Mattice \,\cite{LSM}. The longitudinal spin-spin correlation function is
$$
{\cal K}^{z}(r)-  m_{z}^{2} \sim r^{-2}
$$
and the transverse correlation function decays more slowly 
$$
{\cal K}^{\em tr}(r) \sim r^{-1/2}.
$$
For further results on these correlation functions, see the Ref. \cite{KorepIzegBog}.

In our calculations of correlation functions we follow the route 
used in Ref. \cite{LSM}. In this approach the correlation functions are
given in terms of Toeplitz determinants. Using the Jordan-Wigner 
transformation (\ref{JWtransformation}) one can easily found that
\bea\label{g}
&{\cal K}^{z}(r) = det \, \hat{{\bf g}}=\frac{1}{4}
\left| \begin{array}{cc} 
g(0) & g(r) \\
g(r) & g(0) 
\end{array} \right|\nonumber\\
& = \frac{1}{4} \left(g^{2}(0)-g^{2}(r) \right).&
\eea
and
\bea\label{correelationxxAlphalessap}
&{\cal K}^{\em tr}(r) = det \, \hat{{\bf G}} =&\nonumber\\
\vspace{3.0mm}
&= \frac{1}{4}
\left| \begin{array}{cccc}
g(1) & g(2) & \ldots &  g(r) \\
g(0) & g(1) & \ldots &  g(r-1) \\
\ldots & \ldots & \ldots & \ldots \\
\ldots & \ldots & \ldots & \ldots \\
g(2-r) & g(3-r) & \ldots &  g(1)
\end{array} \right|\, ,&
\eea
Here
\be\label{gr1}
g(r)=\frac{2}{\pi r }~\sin(k_{F}^{-}r) - \delta_{0r}
\ee
at $\alpha<\alpha_{c}$, and 
\be\label{gr2}
g(r)=\frac{2}{\pi r} \left[\sin(k_{F}^{-}r)-
\sin(k_{F}^{+}r)\right] -\delta_{0r}
\ee
at $\alpha > \alpha_{c}$.

After straightforward calculations we obtain that for $\alpha < \alpha_{c}$
\be\label{zzCorreelationAlphaless}
{\cal K}^{z}(r)=  m_{z}^{2}-\frac{1}{\pi^{2}}\frac{\sin^{2}(k^{-}_{F}\cdot r )}{r^{2}}
\ee
and for $\alpha > \alpha_{c}$
\be\label{zzCorreelationAlphamore}
{\cal K}^{z}(r)=  m_{z}^{2}- \frac{1}{\pi^{2}}\frac{\left(\sin(k^{-}_{F} \cdot r  )\nonumber\\
-\sin(k^{+}_{F}\cdot r ) \right)^{2}}{r^{2}}\, .
\ee
As we observe, the quadratic decay of the longitudinal correlations remains 
unchanged at the transition point. However, the additional oscillations in the longitudinal correlation function 
$\sim\cos(2k_{F}^{+}\cdot r)$ and $\sim\cos\left((k_{F}^{-} \pm k_{F}^{+})\cdot r\right)$ 
associated with the presence of two Fermi points are clearly seen.

The transverse correlation undergoes more dramatic change. We obtain
that, for $\alpha<\alpha_{c}$
\be\label{xyCorreelationAlphaless} 
{\cal K}^{\em tr}(r)= \frac{A}{r^{1/2}} +  
\frac{B\cdot \cos(2k_{F}^{-} \cdot r)}{r^{5/2}}
\ee
and for $\alpha > \alpha_{c}$.
\bea\label{xyCorreelationAlphamore}
{\cal K}^{\em tr}(r) &=&\frac{B_{1} \cos(k_{F}^{-} r+\varphi_{-}))}{r}  +
\frac{ C_{1} \cos \left(k_{F}^{+} r+\varphi_{+}\right)}{r}\nonumber\\
&+&\frac{D_{1} \cos\left((2k_{F}^{-} - k_{F}^{+}) r +\varphi_{1}\right)}{r^{3}}\nonumber\\  
&+& \frac{E_{1}\cos\left((2k_{F}^{+} - k_{F}^{-}) r +\varphi_{2}\right)}{r^{3}}\, .
\eea
Here $A=A(\alpha), ... E_{1}(\alpha)$ are the smooth functions of the parameter $\alpha$.

Below the gapless phase with the power-law decay of spin-spin correlations 
given by Eqs.  (\ref{zzCorreelationAlphaless}) and (\ref{xyCorreelationAlphaless}) is called 
the {\em Spin Liquid I} (SL-I) phase, while the gapless phase with the
power-law decay of spin-spin correlations given by
Eqs. (\ref{zzCorreelationAlphamore}) and (\ref{xyCorreelationAlphamore}) - the 
{\em Spin Liquid II} (SL-II) phase. 

The behavior of the correlation functions given by Eqs. 
(\ref{zzCorreelationAlphaless})-(\ref{xyCorreelationAlphamore})
perfectly fits into the usual conformal theory results for the 
systems with several gapless excitations \cite{Frahm,ZKZ}.

\subsection{Emptiness formation probability}

The very important quantity which characterizes a quantum spin system in the 
spin-liquid phase is the so called emptiness formation probability (EFP) 
\cite{KorepinIzedgin}, i.e. the probability to find a ferromagnetic string of the length 
``{\it n}'' in the spin liquid ground state
\be\label{EFP} 
P(n)=\langle GS| \prod_{j=1}^{n} (S^{z}_{j}+\frac{1}{2}) |GS \rangle \, .
\ee
To calculate the EFP we follow the route developed in the recent paper by M. Siroishi, M. Takahashi and 
I. Nishiyama \,\cite{TakahashiEFP}. In the SF representation the EFP is described in terms of the 
following determinant
\bea\label{EFPdeterminant}
&P(n) = \left| \begin{array}{cccc}
f_{11} & f_{12}  & \ldots &  f_{1n} \\
f_{21}& f_{22} & \ldots &   f_{2n}\\
\ldots & \ldots & \ldots & \ldots \\
\ldots & \ldots & \ldots & \ldots \\
f_{n1} & f_{n2} & \ldots &  f_{nn}
\end{array} \right|\, ,&
\eea
where
\bea\label{fnm}
&f_{nm} = \langle GS|c^{\dagger}_{n}c_{m} |GS \rangle&\nonumber\\ 
&= 1/(2\pi) \int_{\Lambda} dq \exp(-iq(n-m)) \, .&
\eea
For $\alpha < \alpha_{c}$ 
\be\label{fmnLess}
f_{nm}=\frac{2}{\pi (n-m)}\sin(k_{F}^{-}(n-m)/2)\cos(k_{F}^{-}(n-m)/2)
\ee
and for $\alpha > \alpha_{c}$ 
\bea\label{fmnMore}
f_{nm}&=&\frac{1}{\pi (n-m)}[\sin(k_{F}^{-}(n-m)) - \sin(k_{F}^{+}(n-m)) ]\nonumber \\
&=& \frac{2}{\pi (n-m)}\sin([k_{F}^{-}-k_{F}^{+}](n-m)/2)\nonumber \\
&\times&\cos([k_{F}^{-}+k_{F}^{+}](n-m)/2)
\eea
\vspace{5mm}
\begin{figure}
\epsfxsize=85mm
\centerline{\epsfbox{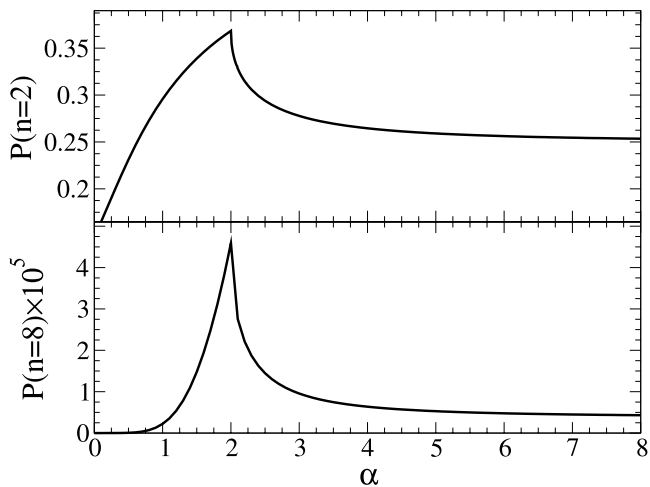} }
\vspace{-2mm}
\caption{The emptiness formation probability $P(n)$ at $n=2$ and $n=8$
as a function of the parameter $\alpha$. }
\label{fig:EFP}  
\end{figure}
\vspace{5mm}

Using Eqs. (\ref{EFPdeterminant})- (\ref{fmnMore}) we calculated the exact values of EFP 
$P(n)$ for $n=1,...,20$. The results of these calculations are presented in the Table 1 
(see Appendix A). 
In Fig. \ref{fig:EFP} we plotted the EFP $P(n)$ as a function of the parameter 
$\alpha$ for $n=2$ and $n=8$. As it is clearly seen from this figure the EFP shows the  
nonmonotonic behavior at $\alpha = \alpha_{c}$.

To consider whether QPT leads to the change in the asymptotic ($n \rightarrow \infty$) 
behavior of EFP we derived the analytical expressions for EFP at $\alpha < \alpha_{c}$ and 
$\alpha > \alpha_{c}$ which fit the numerical data.

At $\alpha < \alpha_{c}$, (in full agreement with the results obtained
in the Ref.\cite{TakahashiEFP} ) we obtain the following expression for EFP 
\be\label{EFP1} 
P_{<}(n)= C \left(\cos\frac{k_{F}^{-}}{2}\right)^{-1/4} n^{-1/4}  \left(\sin\frac{k_{F}^{-}}{2}\right)^{n^{2}}\, ,
\ee
where $C=0.6450$. 

At $\alpha \gg \alpha_{c}$ we obtain the following expression for EFP, which fits our 
exact data
\bea\label{EFP2} 
P_{>>}(n) &=& \left[\sin\left(\frac{k_{F}^{-}- k_{F}^{+}}{2}\right)
\sin\left(\frac{k_{F}^{-}+ k_{F}^{+}}{2}\right)\right]^{\frac{n^{2}}{2}}\nonumber\\
&\times& \left(\frac{A_{1}+(-1)^{n}B_{1}}{n^{1/2}}\right) ,
\eea
where $A_{1}=0.659  $ and $B_{1}=0.054$. In the intermediate regime,
for $\alpha \geq \alpha_{c} $ the fitting formula is
more complicated and is presented in Appendix B. As we see, the QPT in the system manifestly 
changes the EFP.

\subsection{Order parameter}

The QPT at $\alpha=\alpha_{c}$ from the SL-I to the SL-II 
phase requires an Order Parameter for complete description. As the order parameter we introduce 
the following quantity                                           
\begin{equation} \label{OrderParameter1}
\eta(\alpha)=\bar{{\cal L}}_{\small ExtXY} - \bar{{\cal L}}_{\small SL-I}
\end{equation}  
where
\begin{equation} \label{barlGXY}
\bar{{\cal L}}_{\small ExtXY}  = \frac{\sum  n \cdot  P(n)}{\sum P(n)}
\end{equation}  
is the average length of the ferromagnetic string in the ground state of the extended $XY$ 
chain with coupling $\alpha$ and, respectively, the magnetization $m^{z}(\alpha)$, while 
$\bar{{\cal L}}_{\small SL-I}$ is the same quantity calculated in the SL-I 
phases with the same value of the magnetization.  
\vspace{5mm}
\begin{figure}
\epsfxsize=85mm
\centerline{\epsfbox{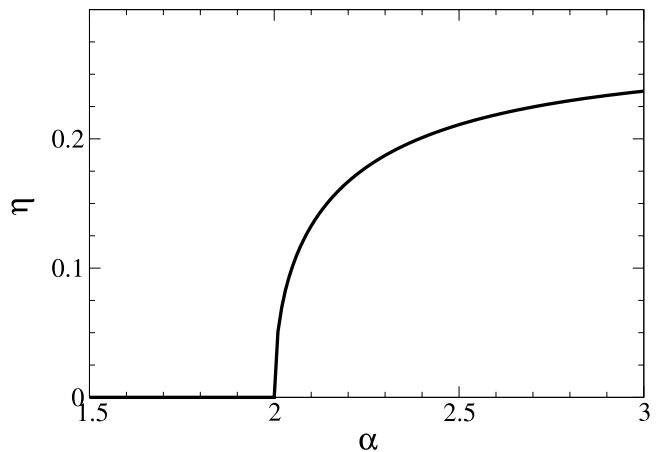} }
\vspace{-2mm}
\caption{Order Parameter as a function of the parameter $\alpha$.}
\label{fig:OrderparameterEFP}  
\end{figure}
\vspace{5mm}

Straightforward calculations give (see Fig. \ref{fig:OrderparameterEFP} ) 
\begin{equation} \label{OrderParameter2}
\eta(\alpha)=\left\{
\begin{array}{l}
{\em 0} \hskip1.7cm {\mbox at} 
\hskip0.3cm (\alpha-\alpha_{c}\rightarrow 0_{-}) \\
\sqrt{\alpha-\alpha_{c}} \hskip0.8cm {\mbox at}
\hskip0.1cm  (\alpha-\alpha_{c}) \rightarrow 0_{+})
\end{array} 
\right. \, . 
\end{equation}  
Thus, the order parameter exhibits the standard mean-field type behavior. 

\section{Thermodynamics}

The QPT with respect to the parameter $\alpha$ in the ground-state of
the system (\ref{GXYhamiltonian}) is smeared
out by thermal fluctuations at finite temperatures. Thus, the thermodynamic 
quantities have no singularities but, in the low-temperature region 
$T \ll J $ they show a well pronounced anomalous behavior at 
$|\alpha - \alpha_{c}| \ll T/J$.

To investigate the thermodynamics of the system let us first calculate the
free energy. In the thermodynamic limit the free energy of the system
is given by
\be\label{FreeEnergu}
F= -T\int d\omega\rho(\omega)
\ln\left[2\cosh\left(\omega/2T\right)\right]\, .
\ee
The SF density of states $\rho(\omega)$ is nonzero and equal to
\begin{equation}\label{rho1}
\rho(\omega) =\rho^{-}(\omega)
\end{equation}
for $\alpha < 2$ and for $\alpha > 2$ and $E(0) < \omega <
W_{max}$. For $\alpha > 2$ and $W_{min} < \omega < E(0)$   
\begin{equation}\label{rho2}
\rho(\omega) =\rho^{-}(\omega)+\rho^{+}(\omega)\, .
\end{equation}
Here $W_{min}={\em min}\{-J(2-\alpha)/2,-J(2\alpha^{2} +1)/4\alpha\}$, $W_{max}=  J(\alpha+2)/2$, and
\begin{equation}\label{ZZcorrelatorXY}
\rho^{\pm}(\omega)= \frac{1}{\pi J}\frac{2\alpha} {{\cal G}_{\alpha}(\omega/J)}
\frac{1}{\sqrt{4\alpha^{2} - \left [1 \pm {\cal G}_{\alpha}(\omega/J) \right]^{2}}}\, ,
\end{equation} 
where
\begin{equation}
{\cal G}_{\alpha}(\omega/J)=\sqrt{1+2\alpha^{2} + 4\alpha(\omega/J)}\, .
\end{equation} 

Knowing the free energy it is easy to obtain the expression 
for the entropy
\bea\label{ZZcorrelatorXY3}
&S = \int^{\infty}_{-\infty}d\omega\rho(\omega) \Big[\ln\left(2\cosh\left(\frac{\omega}{2T}\right)\right)&
\nonumber\\ 
&-\left(\frac{\omega}{2T}\right)\tanh\left(\frac{\omega}{2T} \right)\Big].&
\eea

Below in this Section we study the anomalies in the low-temperature behavior of different thermodynamic 
quantities, caused by the presence of the QPT in the ground state at $\alpha \simeq \alpha_{c}$.

\subsection{Specific heat}

The anomalous temperature dependence in the vicinity of the QPT is seen most clearly  in the 
low-temperature behavior of the specific heat. The heat capacity of the system is given by
\begin{equation}\label{ZZcorrelatorXY1}
C=\int^{\infty}_{-\infty}d\omega\rho(\omega)
\frac{(\omega/2T)^{2}}{\cosh^{2}\left(\omega/2T\right)}
\end{equation} 

Away from the critical point, the density of states in the vicinity of the
Fermi level $\rho^{-}(0)$ at $\alpha \ll \alpha_{c}$  and $\rho^{-}(0)+\rho^{+}(0)$ 
at $\alpha \gg \alpha_{c}$, respectively, is the constant of the order of $\sim 1/J$. 
Therefore, in these cases the specific heat of the system
$$
C(T) \simeq \gamma \left(T/J\right)
$$
with $\gamma \sim 1$. 

However, in the vicinity of the critical point, at 
$\left|\alpha-\alpha_{c}\right| \leq \left|\omega/J\right|$,
\begin{equation}\label{ZZcorrelatorXY4}
\rho^{+}(\omega)\simeq \frac{1}{\pi\sqrt{6J|\omega| }}
\end{equation} 
and therefore, in this case the heat capacity of the system exhibits 
the anomalous square root dependence on the temperature 
\begin{equation}\label{ZZcorrelatorXY5}
C= \gamma_{1}\cdot \sqrt{T/J}  + \gamma \cdot (T/J).
\end{equation} 
\begin{figure}
\epsfxsize=85mm
\centerline{\epsfbox{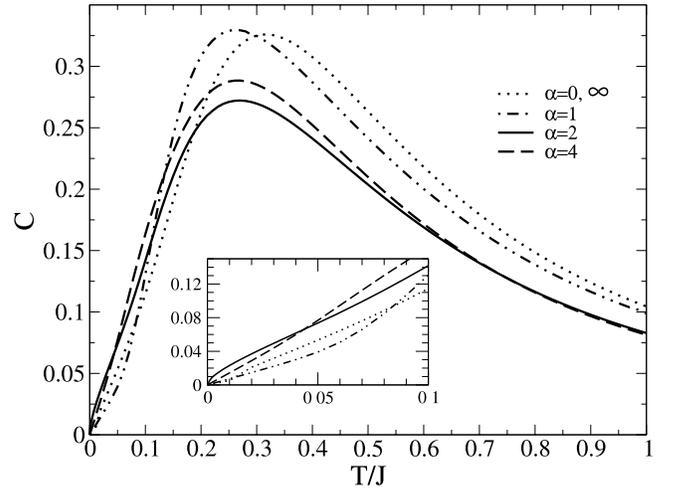} }
\vspace{-2mm}
\caption{The specific heat of the system as a function of the 
parameter $T/J$ for different values of the parameter $\alpha$. The inset 
shows the specific heat of the system at $T/J \ll 1$. }
\label{fig:CTinCT}
\end{figure}
\vspace{5mm}

In Fig. \ref{fig:CTinCT} we plotted the heat capacity of the extended $XY$ model 
for different values of the parameter $\alpha$. Since the bandwidth of the system depends on 
the parameter $\alpha$, in Fig. \ref{fig:CTinCT} we plotted the results for the models with 
normalized bandwidth $W_{R}=2J$. In the inset we present the low-temperature 
$(T\ll J$) heat capacity. The anomaly in the behavior of heat capacity at 
$\alpha=\alpha_{c}=2$ is clearly seen.

\subsection{Magnetization and magnetic susceptibility at $T\neq0$.}

At finite temperature the magnetization of the system is proportional to the average number of 
spinless fermions
\be\label{Magnetization2} 
m_{\alpha}^{z}(T)=  -\frac{1}{2}\int_{-\infty}^{\infty} d\omega\rho(\omega)\tanh\left(\omega/2T\right)\, , 
\ee 
while the magnetic susceptibility is given by 
\be
\chi(T) =\frac{1}{4T} \int_{-\infty}^{\infty} 
d\omega \rho(\omega)\frac{1}{\cosh^{2} \left(\omega/2T \right)}\, .
\ee

Let us first consider the limiting case of small temperatures $T/J \ll 1$. In this case 
it can be easily obtained, that for $\alpha \ll \alpha_{c} $
\bea\label{magnetizzationTSmallalpha}
m_{\alpha}^{z}(T) &=& m_{\alpha}^{z}(0)\left(1-\frac{\pi^{2}}{2} \cdot (T/J)^{2}
+ {\cal O}((T/J)^{4})\right)\, , \\
\chi_{\alpha}(T) &=& \chi_{\alpha}(0) \left(1+\frac{\pi^{2}}{12} \cdot (T/J)^{2} + {\cal O}((T/J)^{4})\right)\, ,
\eea
and for $\alpha \gg \alpha_{c} $
\bea\label{magnetizzationTLargealpha}
m_{\alpha}^{z}(T) &=& m_{\alpha}^{z}(0) \left( 1 + 2\pi^{2} \cdot (T/J)^{2} + {\cal O}((T/J)^{4})
\right)\, , \\
\chi_{\alpha}(T) &=&\chi_{\alpha}(0) \left(1 + \frac{2\pi^{2}}{3} \cdot (T/J)^{2} + {\cal O}((T/J)^{4}) \right)\, .
\eea
\begin{itemize}
\item{The  low temperature ($T/J \ll 1$) expansion shows that magnetization of the
system decays with temperature at $\alpha < \alpha_{c}$, and {\em magnetization
of the system increase with the increase of the temperature} at $\alpha > \alpha_{c}$.}
\item{The magnetic susceptibility of the system is finite both for $\alpha \ll \alpha_{c}$ and
$\alpha \gg \alpha_{c}$ and shows the inverse square root dependence on temperature for
$\left|\alpha - \alpha_{c}\right| \ll T/J$.}
\end{itemize}

In the case of coupling close to the critical value, at $\left|\alpha - \alpha_{c}\right| \ll T/J$ 
\bea\label{magnetizzationTalphaCritical} 
m_{\alpha}^{z}(T) &=& m_{\alpha}^{z}(0) 
\Big[ 1 - \beta_{1/2} \cdot (T/J)^{1/2} - \beta_{3/2}\cdot(T/J)^{3/2}\nonumber\\ 
&-& \beta_{2}\cdot(T/J)^{2} + {\cal O}((T/J)^{5/2})\Big]\, ,\\
\chi(T) & =&\beta_{-1/2}\cdot \left(TJ \right)^{-1/2}+\rho_{\alpha}^{-}(\omega=0) + {\cal O}(T/J)^{2}\, .
\eea
Here
\bea\label{B12} 
\beta_{1/2} &=& -\frac{\sqrt{2}-1}{\sqrt{12\pi}}\zeta(1/2) > 0\, , \\
\beta_{3/2} &= &\frac{\sqrt{2}-1}{96\sqrt{6\pi}}\zeta(3/2) > 0\, , \\
\beta_{2} &= &\frac{\pi}{81\sqrt{3}}\, , \\
\beta_{-1/2}&= &\frac{1}{\sqrt{6\pi}} ( 2\sqrt{2}-1) \zeta(-1/2) > 0\, ,
\eea
and  $\zeta(z)$ is the Riemann's zeta function.

\vspace{5mm}
\begin{figure}
\epsfxsize=85mm
\centerline{\epsfbox{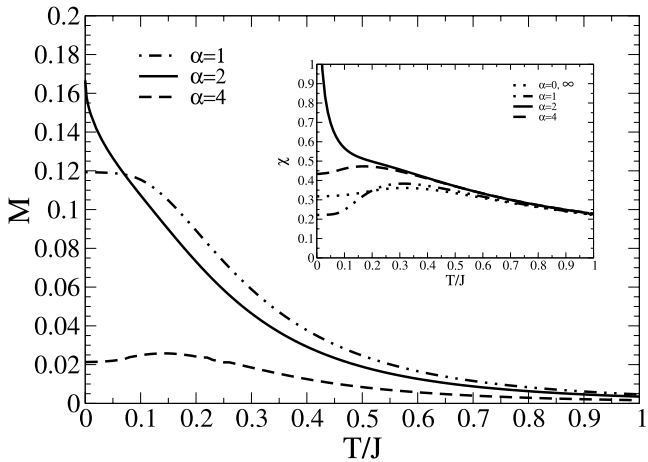} }
\vspace{2mm}
\caption{Magnetization of the system as a function of temperature for 
different values of the parameter $\alpha$. The inset shows the
magnetic susceptibility of the system as a function 
of temperature for different values of the parameter $\alpha$.}
\label{fig:MagnetizationT}
\end{figure}
\vspace{2mm}

Let us now consider in detail the problem of {\em anomalous temperature dependence} of magnetization for 
$\alpha > \alpha_{c}$. In Fig. \ref{fig:MagnetizationT} we plotted the  
magnetization of the system as a function of temperature for three different values of the parameter 
$\alpha = 1, 2$ and $4$. From Fig. \ref{fig:MagnetizationT} it is clearly
seen that at $\alpha=4$ the magnetization reaches its maximum at $T=0.15J$ and then smoothly decays with the 
increase of temperature.

\begin{figure}
\epsfxsize=85mm
\centerline{\epsfbox{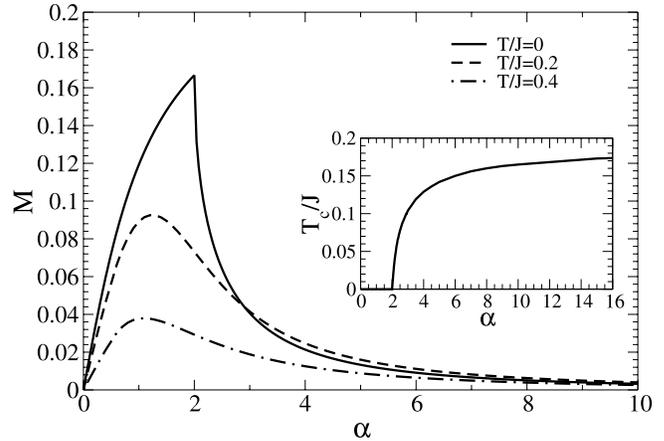} }
\vspace{5mm}
\caption{Magnetization of the system as a function of the parameter $\alpha$ 
at different values of temperature. } 
\label{fig:MAlphaT}
\end{figure}

To gain more insight into the behavior of magnetization in Fig. \ref{fig:MAlphaT} we plotted 
the magnetization as a function of the parameter $\alpha$ for different values of 
the temperature.  As it is clearly seen from Fig. \ref{fig:MAlphaT},  $M(T) < M_{0} \equiv M(T=0)$ at 
arbitrary $T$ when $\alpha < \alpha_{c}$. 
But at $\alpha > \alpha_{c}$ there does exist the {\em crossover temperature} $T_{c}$, and so  
$M(T) > M_0$ for $0 < T < T_{c}$ and  $M(T) < M_0$  for  $T > T_{c}$. We calculated 
$T_{c}$ as a function of the 
parameter $\alpha$ (see inset in Fig. \ref{fig:MAlphaT}) and found that $T_{c}$ monotonically increases 
with the increase of  $\alpha$ from its minimum value $T_{c}=0$ at $\alpha=\alpha_{c}=2$ up to 
$T_{c}=0.186J$ at $\alpha \rightarrow \infty$ .

\vspace{3mm}
\vspace{3mm}

The source of such unusual behavior is the structure of the excitation spectrum: at $\alpha \neq 0$ 
the ground state of the system is characterized by the magnetization $M_{0}^{\alpha}$. 
At $\alpha < \alpha_{c}$, the excited states with magnetization $M =
M_{0}^{\alpha} \pm m$ have equal energy for arbitrary
$m=1,2,...$. Therefore, for $\alpha < \alpha_{c}$ at arbitrary finite temperature 
$M^{\alpha}(T) < M_{0}^{\alpha}$. On the other hand, for
$\alpha > \alpha_{c}$ the excitation spectrum of the system exhibits a clear anisotropy: 
at least some lowest excited states with magnetization $M=M_{0}^{\alpha} + m $ (with $m=1,2...,m_{0}$) are far 
below their counterparts with magnetization $M=M_{0}^{\alpha} - m $. Therefore, at sufficiently 
low temperatures, where the states with magnetization $M > M^{0}_{\alpha} $ are occupied 
mainly, the total magnetization of the system increases. However, with further increase 
of the temperature, the difference in occupation of states with
$M=M^{0}_{\alpha}  \pm m$ reduces, and 
finally, at $T> T_{c}$, the magnetization of the system drops below its ground state value.

\section{Properties of the system in the presence of external magnetic field}

In this Section we consider the properties of the extended $XY$ spin chain 
(\ref{GXYhamiltonian}) in the presence of an external magnetic field. We consider the 
general case, where the field dependent part of the Hamiltonian is given by
\begin{equation} \label{GXYMagFieldHamiltonian}
{\cal H}= -  \sum_{n}\left(H_{0} + (-1)^{n} H_{\delta} \right) S^{z}_{n} \, . 
\end{equation} 

In terms of spinless Fermions the Hamiltonian of the extended XY model in the presence 
of an external magnetic field reads
\bea \label{GXYSFhamiltonianMagField} 
{\cal H} &=& -\frac{J}{2}\sum_{n}\left(c_{n}^{\dagger}c_{n+1} + 
c^{\dagger}_{n+1}c_{n}\right) \nonumber\\
&+& \frac{J^{\ast}}{4}\sum_{n}\left(c_{n}^{\dagger}c_{n+2} +
c^{\dagger}_{n+2}c_{n}\right)\nonumber\\
 &- & \sum_{n} \left(H_{0}+(-1)^{n}H_{\delta} \right) 
\left( c^{\dagger}_{n}c_{n}-\frac{1}{2}\right)\, .
\eea

Below in this Section we consider separately the ground state phase diagram of the 
extended $XY$ spin chain in the case of:

\begin{itemize}
\item{applied uniform magnetic field $H_{\delta}=0$;}
\item{applied staggered magnetic field $H_{0}=0$;}
\item{applied magnetic field acting only on the spins on the even sites $H_{0}=H_{\delta}=H$.}
\end{itemize}

\subsection{Uniform magnetic field}

In the case of applied uniform magnetic field ($H_{\delta}=0, \, H_{0}=hJ $) 
the diagonalization of the Hamiltonian is straightforward and gives
\begin{equation} \label{GXYSFhamiltonianUnifH}
{\cal H}=\sum_{k}E_{1}(k)c^{\dagger}(k)c(k)
\end{equation} 
where
\begin{equation} \label{Spectrum1}
E_{1}(k)=-J\left(h + \cos(k)-\frac{\alpha}{2}\cos2k\right) \, .
\end{equation} 

\begin{figure}
\epsfxsize=85mm
\centerline{\epsfbox{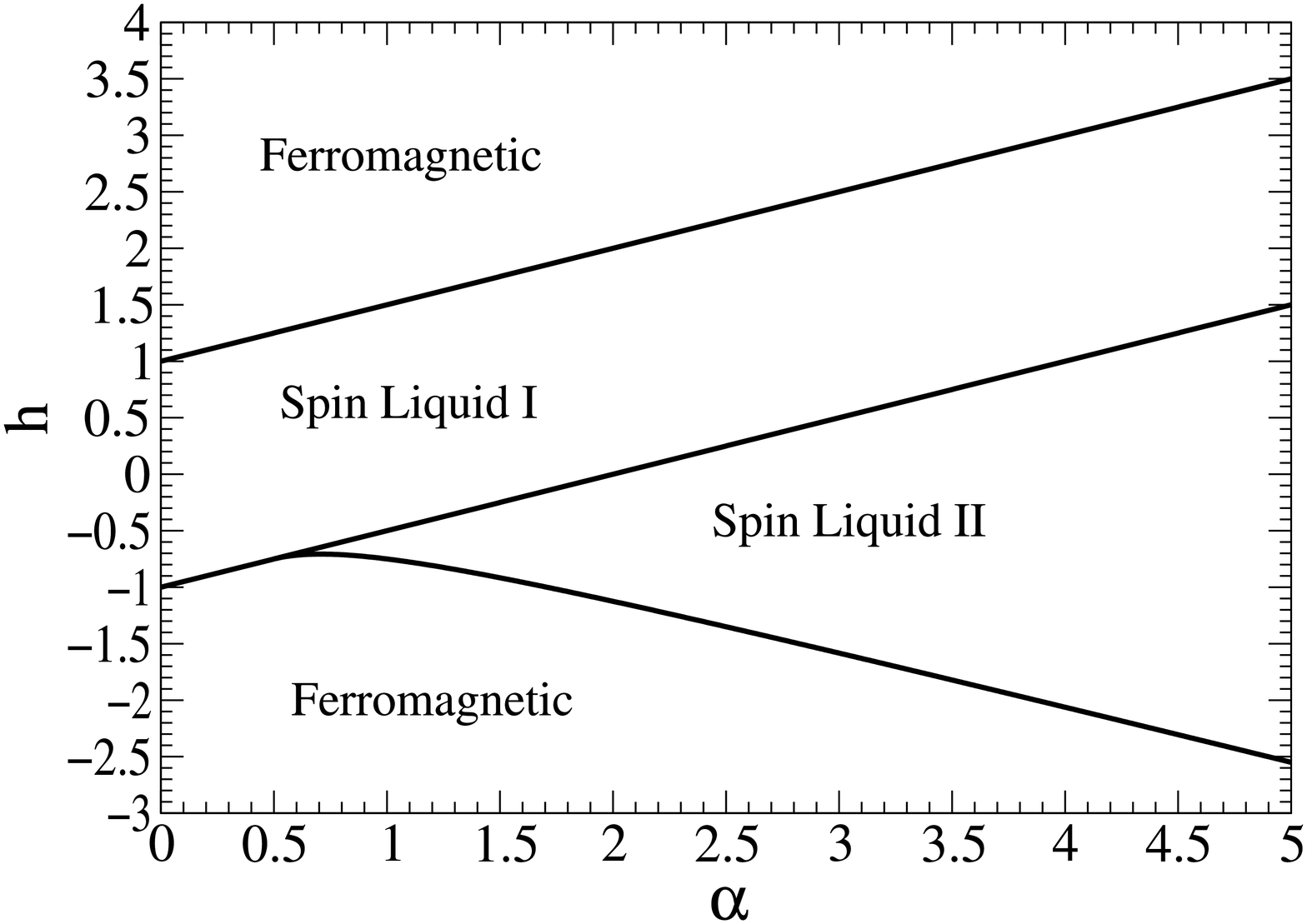}} 
\vspace{2mm}
\caption{The ground state phase diagram of the model in the case of applied uniform magnetic field.}
\label{fig:PhDiagUnMagField}
\end{figure}

The phase diagram of the model in the case of applied uniform magnetic field is  
presented in Fig.\ref{fig:PhDiagUnMagField}: 
\begin{itemize}
\item{For $hJ < W_{min} $ and  $hJ > W_{max}$ the system is in the 
{\em ferromagnetic phase }}.
\item{For $- \left(1-\frac{1}{2}\alpha \right) < h <  \left(1 +\frac{1}{2} \alpha\right)$ the 
system  is in the {\em SL-I} phase.}
\item{For $\alpha > 0.5 $ and $- \left(2\alpha^{2} + 1\right)/4 \alpha < h  < - \left( 1 - 
\frac{1}{2}\alpha  \right)$ the system is in the {\em SL-II} phase.}
\end{itemize}
\vspace{5mm}
\vspace{5mm}

\subsection{Staggered magnetic field}

In the case of applied staggered magnetic field ($H_{0}=Jh,  H_{\delta}=Jh_{\delta}$) it is useful to introduce 
two types of spinless fermions determined on the odd and even sub-lattices, respectively
\begin{equation} \label{ab} 
c_{2n-1} \equiv a_{n-1/2} \qquad {\mbox and}\qquad c_{2n}\equiv b_{n}.
\end{equation}  

In terms of these new particles, the Hamiltonian in the momentum space reads
\bea \label{GXYSFhamiltonianhohd} 
{\cal H}& = & \sum_{k} \Big[( \varepsilon_{a}(k) a^{\dagger}_{k}a_{k}+
\varepsilon_{b}(k) b^{\dagger}_{k}b_{k} \nonumber\\
&+& \varepsilon_{ab}(k) \left(a^{\dagger}_{k}b_{k}+b^{\dagger}_{k}a_{k} \right) \Big] 
\eea
where
\bea \label{varepsilons}
\varepsilon_{a}(k) &=& \frac{J^{\ast}}{2}\cos(2k)-h_{0}+h_{\delta}\, ,\nonumber\\
\varepsilon_{b}(k)&=&\frac{J^{\ast}}{2}\cos(2k)-h_{0}-h_{\delta} \, ,\\
\varepsilon_{ab}(k) &=&-J \cos k\, .\nonumber 
\eea
Using the standard Bogoliubov rotation
\begin{eqnarray} \label{Bogolubov} 
a_{k}=\cos \left(\frac{1}{2} \theta_{k} \right) \alpha_{k}+    
\sin\left(\frac{1}{2} \theta_{k} \right) \beta_{k}  \nonumber\\  
b_{k}=-\sin \left(\frac{1}{2} \theta_{k} \right) \alpha_{k}+    
\cos \left(\frac{1}{2} \theta_{k} \right) \beta_{k} 
\end{eqnarray} 
where
\begin{equation} \label{delta}
\tan(\theta_{k})= \frac{2\varepsilon_{ab}(k)}{\varepsilon_{a}(k)-\varepsilon_{b}(k)}  
\end{equation} 
we finally obtain
\be
{\cal H} = \sum_{k}\left( E_{-}(k)\alpha^{\dagger}_{k} \alpha_{k} + 
E_{+}(k)\beta^{\dagger}_{k}\beta_{k} \right) \, ,
\ee 
where
\be \label{speqtrumStaggaredField} 
E_{\pm}(k)= J \left( \frac{\alpha}{2}\cos2k - h \pm \sqrt{h_{\delta}^{2}+ \cos^{2}k} \right) \, . 
\ee

\begin{figure}
\epsfxsize=85mm
\centerline{\epsfbox{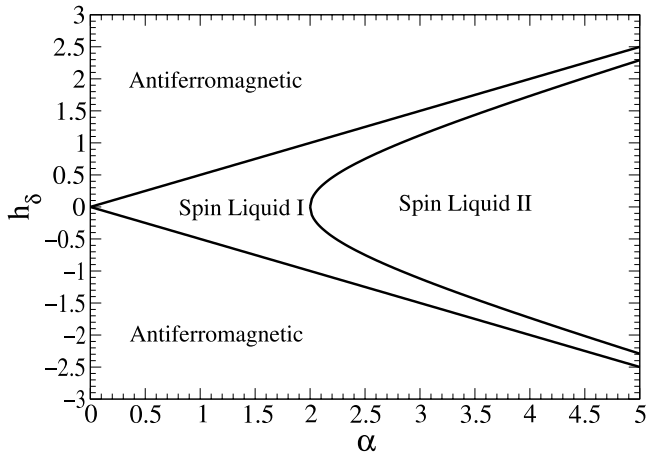} }
\vspace{.2in}
\caption{The ground state phase diagram of the model in the case 
of applied staggered magnetic field.}
\label{fig:PhDiagAlterMagField}
\end{figure}

Let us first consider the case of the staggered magnetic field  
($H_{0}=0$).  The ground state phase diagram of the model in this case 
consists of the following four sectors (see Fig.\ref{fig:PhDiagAlterMagField} ): 
\begin{itemize}
\item{At $\left |h_{\delta} \right|  > \frac{1}{2} \alpha$ the system is in the 
long-range ordered Ne\'el {\em anti-ferromagnetic phase }.} 
\end{itemize}
There is a gap in the spin 
excitation spectrum. The longitudinal spin-spin correlation function is 
$$
{\cal K}_{z}(r) = \langle S_{n}^{z}S_{n+r}^{z} \rangle \simeq (-1)^{r} f(\alpha)\, 
$$
where $f(\alpha)$ is the constant of the order of unity for given $\alpha$.
\begin{itemize}
\item{At $\alpha < 2 $ and $\left |h_{\delta} \right|  < \frac{1}{2} \alpha$ as well 
as for $\alpha > 2$ and  $\sqrt{\frac{\alpha^{2}-4}{4}} <\left |h_{\delta}\right| 
< \frac{1}{2} \alpha$
the system is in the {\em SL-I} phase}
\item{At $\alpha > 2$ and  $\left|h_{\delta}\right| < \sqrt{\frac{\alpha^{2}-4}{4}}$
the system is in the {\em SL-II} phase.}
\end{itemize}

\vspace{.2in}

\subsection{Mixed magnetic field}

In this subsection we consider a rather special case, where the mixed magnetic field with equal strength of the 
uniform and staggered components $H_{0}=H_{\delta}=hJ$. This corresponds to the case, where the magnetic field 
of the $2hJ$ strength is applied to the spins on the even sites, while the spins on the 
odd sites experience no magnetic field.

The spectrum of the system in this case is given by
\begin{equation} \label{Spectrum1a}
\tilde{E}^{\pm}_{2}(k)=J \left(\frac{\alpha}{2}\cos2k-h \pm  \sqrt{h^{2} + \cos^{2}k}\right) \, .
\end{equation} 
\vspace{5mm}
\begin{figure}
\epsfxsize=85mm
\centerline{\epsfbox{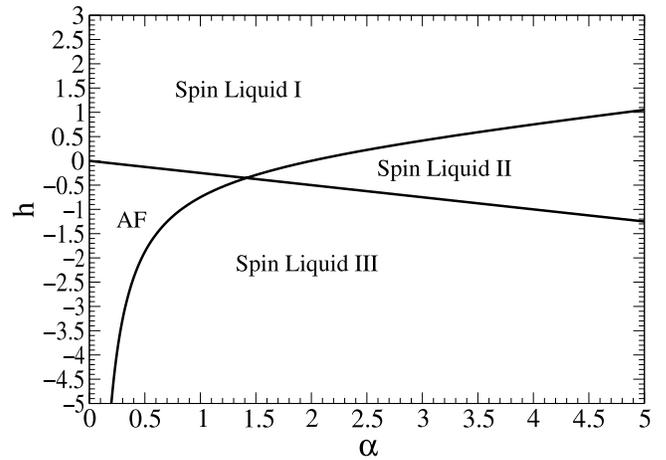} }
\vspace{-3mm}
\caption{Phase diagram for Uniform Magnetic field + alternating Magnetic
field } 
\label{fig:PDmixedMagneticFeld}
\end{figure}
\vspace{3mm}

The ground state phase diagram of the model (\ref{GXYSFhamiltonianhohd}) in the case of applied 
mixed magnetic field consists of the following four sectors 
(see Fig.\ref{fig:PDmixedMagneticFeld}) 
\begin{itemize}
\item{At $\alpha < \sqrt{2}$ and $-\frac{4-\alpha^{2}}{4\alpha} < h < 
- \frac{1}{4} \alpha$ the system is in the long-range ordered Ne\'el 
{\em antiferromagnetic phase };} 
\item{At $\alpha > \sqrt{2}$ and  $ -\frac{1}{4} \alpha < h < 
-\frac{4-\alpha^{2}}{4\alpha}$ the system is in the {\em SL-II} phase;}
\item{At  $h > {\mbox max}\{-\frac{1}{4} \alpha; -\frac{4-\alpha^{2}}{4\alpha}\}$ the 
system is in the {\em SL-I} phase;}  
\item{At $h < {\mbox min}\{-\frac{1}{4} \alpha; -\frac{4-\alpha^{2}}{4\alpha}\}$  the 
system is in the {\em SL-III} phase;} 
\end{itemize}

The new {\em SL-III} phase corresponds to the gapless phase characterized by  
qualitatively different behavior of the long-range spin-spin correlations for the spins 
located on the even and odd sites. The calculation of the longitudinal
spin-spin correlations between the spins on the even sites gives
\be\label{zzCorrelatorsEE}
{\cal K}_{0, 2m}^{z}(r=2m) = (m_{z}^{even})^{2}\, ,
\ee
between the spins on the odd sites 
\be\label{zzCorrelatorsOO} 
{\cal K}_{1, 2m+1}^{z}(r=2m) = (m_{z}^{odd})^{2} -\frac{B}{r^{2}}, 
\ee
and between the spins on even and odd sites
\be\label{zzCorrelatorsOE}
{\cal K}_{1, 2m}^{z}(r=2m-1)= m_{z}^{even}m_{z}^{odd}
\ee
respectively. Here $m_{z}^{even}< 0$ and $m_{z}^{odd}> 0$ is the magnetization per site of the even and odd 
sub-lattices, respectively, which is calculated at the given value of the coupling $\alpha$. The parameter $B$ is a 
smooth function of the coupling constant $\alpha$ and at the given $\alpha$ it is the constant of the order of 
unity. For example, at $ h=-0.4$ and  $ \alpha=1.5$ $m_{z}^{even}=-0.3406$ and $m_{z}^{odd}=0.1690$. 
Therefore, in the SL-III phase, 
we have antiferromagnetically coupled two ferromagnetically ordered sub-lattices.

 In the SL-III phase the transverse correlations also show the unusual
 behavior. In particular, the transverse spin  
correlation between the spins on the even sites is
\be\label{XYCorrelatorsEE}
{\cal K}_{0, 2m}^{\em tr}(r=2m) = 0\, ,
\ee
between the spin on the odd sites 
\be\label{XYCorrelatorsOO} 
{\cal K}_{1, 2m+1}^{\em tr}(r=2m) \simeq\frac{(-1)^{m}}{r^{1/2}},
\ee
and between the spins on the even and odd sites 
\be\label{xyCorrelatorsOE} 
{\cal K}^{\em tr}_{1, 2m}(r=2m-1) \simeq \frac{(-1)^{m}}{r^{3/2}}\, .
\ee

\section{conclusion}

In conclusion, we investigated the ground state phase diagram and the
low-temperature thermodynamics of the extended $XY$ model with nearest-neighbor
exchange ($J$) and three-spin interaction $J^{\ast}=\alpha J$. At $J^{\ast}\neq 0$ the spin system
is characterized by the broken time reversal symmetry and for arbitrary
$\alpha \neq 0,\infty$ shows a finite magnetization in the ground
state $M^{0}\left(\alpha \right)$.
We have shown that with the increase of the three-spin coupling the Quantum Phase 
Transition from the Spin-Liquid-I into the Spin-Liquid-II phase takes place. 
In the Spin-Liquid-I phase, at  $\alpha < \alpha^{\ast}_{c}=2$ the properties of the
extended $XY$ chain are similar to those of the standard $XY$ model in the
presence  of such an ``effective'' magnetic field that ensures the same value of
the GS  magnetization $M^{0}$. However, in the Spin-Liquid-II phase, at
$\alpha > \alpha_{c}$, the behavior of the magnetization, magnetic
susceptibility, emptiness formation probability and different
spin-spin correlation functions is qualitatively different and could
not be considered by the simple effect of an ``effective'' magnetic field.  
The anomalous behavior of the heat capacity of the system, 
entropy, magnetization and magnetic susceptibility connected with the
presence of QPT in the ground state has been shown. The ground state
phase diagram of the model in the case of applied uniform and/or
staggered magnetic field has been also obtained.

\section{Acknowledgments}

This work was supported by the SCOPES grant N 7GEPJ62379.
G.I.J. also thanks Arno P. Kampf for hospitality and interesting
discussions during his stay at Augsburg University, where a part of this work has been
done. I.T. also acknowledges the World Federation of Scientists for support.



\newpage

\section{Appendix A}

\end{multicols}

\vspace{5mm} 

In this appendix we present the exact values of the EFP $P(n)$ for 
different values of the parameter 
$\alpha$, obtained by evoluation of the Toeplitz determinant 
(\ref{EFPdeterminant}).
\vspace{10mm} 

\begin{table}
\caption{The emptiness formation probability $P(n)$.}
\begin{center}
\begin{tabular}{|c||c|c|c|c|c|}
\tableline 
\hspace{1mm}& & & &  &\hspace{1mm}  \\
&$\alpha = 0$  & $\alpha = 1$ &$\alpha = 2$ & $\alpha = 4$ & $\alpha = \infty$ \\
\hspace{1mm}& & & &  &\hspace{1mm}  \\
& $\langle m^{z} \rangle = 0$  &  $ \langle  m^{z} \rangle = 0.1193 $ 
& $\langle m^{z} \rangle = 1/6  $ & $\langle  m^{z} \rangle = 0.0213387 $
&$\langle m^{z} \rangle = 0$  \\
\hspace{1mm}& & & &  &\hspace{1mm}  \\
\tableline 
\tableline
\hspace{1mm}& & & & & \hspace{1mm} \\ 
\hspace{1mm} $ n=1 $ & 0.5 & 6.1928 $\cdot 10^{-1} $ & 6.6667 $\cdot
10^{-1} $ & 5.2134 $\cdot 10^{-1}$ & 0.5 \\ 
\hspace{1mm}& & & &  &\hspace{1mm}  \\
\hspace{1mm} $ n=2 $ & 1.4868 $\cdot 10^{-1} $  &   2.9576 $\cdot 10^{-1} $  &   3.6845 $\cdot 10^{-1} $ 
& 2.6455 $\cdot 10^{-1} $&  0.25  \\ 
\hspace{1mm}& & & &  &\hspace{1mm}  \\
\hspace{1mm} $ n=3 $ & 2.3679 $\cdot 10^{-2}$  &1.0251 $\cdot 10^{-1} $&1.6136 $\cdot 10^{-1} $& 
 8.4078 $\cdot 10^{-2}$ & 7.4339 $\cdot 10^{-2}$ \\ 
\hspace{1mm}& & & &  &\hspace{1mm}  \\
\hspace{1mm} $ n=4$ &1.9453 $\cdot 10^{-3}$  &2.5041 $\cdot 10^{-2}$ & 5.4640 $\cdot 10^{-2}$  & 
2.5783 $\cdot 10^{-2}$& 2.2105 $\cdot 10^{-2}$ \hspace{1mm} \\ 
\hspace{1mm}& & & &  &\hspace{1mm}  \\
\hspace{1mm} $ n=5 $  & 8.1263 $\cdot 10^{-5}$  & 4.2525 $\cdot 10^{-3}$ & 1.4127 $\cdot 10^{-2}$  & 
4.5135 $\cdot 10^{-3} $ & 3.5205 $\cdot 10^{-3} $\hspace{1mm} \\ 
\hspace{1mm}& & & &  &\hspace{1mm}  \\
\hspace{1mm} $ n=6$  & 1.7152 $\cdot 10^{-6} $&4.9869 $\cdot 10^{-4}$ &2.7702 $\cdot 10^{-3} $ & 
7.6228 $\cdot 10^{-4}$  & 5.6069 $\cdot 10^{-4}$\hspace{1mm} \\ 
\hspace{1mm}& & & &  &\hspace{1mm}  \\ 
\hspace{1mm} $ n=7$ & 1.8232 $\cdot 10^{-8}$ &4.0241 $\cdot 10^{-5}$  & 4.1052 $\cdot 10^{-4}$  & 
 7.0945 $\cdot 10^{-5}$ &4.6062 $\cdot 10^{-5}$ \hspace{1mm} \\
 \hspace{1mm}& & & &  &\hspace{1mm}  \\
\hspace{1mm} $ n=8$ &9.7402 $\cdot 10^{-11}$ & 2.2298 $\cdot 10^{-6} $ &4.5877 $\cdot 10^{-5}$ & 
6.3746 $\cdot 10^{-6} $& 3.7841 $\cdot 10^{-6} $ \hspace{1mm} \\ 
\hspace{1mm}& & & &  &\hspace{1mm}  \\
\hspace{1mm} $ n=9$ & 2.6122 $\cdot 10^{-13}$& 8.4729 $\cdot 10^{-8}$  &3.8609 $\cdot 10^{-6} $ & 
3.1110 $\cdot 10^{-7} $ & 1.5808 $\cdot 10^{-7} $\hspace{1mm} \\ 
\hspace{1mm}& & & &  &\hspace{1mm}  \\
\hspace{1mm} $ n=10 $ & 3.5137 $\cdot 10^{-16}$ & 2.2060 $\cdot 10^{-9}$  &2.4448 $\cdot 10^{-7} $ & 
1.4664 $\cdot 10^{-8}$ &  6.6036 $\cdot 10^{-9}$ \hspace{1mm} \\
 \hspace{1mm}& & & &  &\hspace{1mm}  \\
\hspace{1mm} $ n=11 $ &2.3691 $\cdot 10^{-19}$ &3.9329 $\cdot 10^{-11}$  &1.1641 $\cdot 10^{-8}$ & 
 3.7291 $\cdot 10^{-10}$ & 1.3938 $\cdot 10^{-10}$\hspace{1mm} \\ 
\hspace{1mm}& & & &  &\hspace{1mm}  \\
\hspace{1mm} $ n=12 $ &8.0036 $\cdot 10^{-23}$  & 4.7992 $\cdot 10^{-13}$&4.1656 $\cdot 10^{-10}$  & 
9.1615 $\cdot 10^{-12}$ & 2.9420 $\cdot 10^{-12}$ \hspace{1mm} \\
 \hspace{1mm}& & & &  &\hspace{1mm}  \\
\hspace{1mm} $ n=13 $ &1.3543 $\cdot 10^{-26}$ &4.0070 $\cdot 10^{-15}$ &1.1200 $\cdot 10^{-11}$ & 
1.2099 $\cdot 10^{-13}$ &3.1272 $\cdot 10^{-14} $ \hspace{1mm} \\ 
\hspace{1mm}& & & &  &\hspace{1mm}  \\
\hspace{1mm} $ n= 14 $ &1.1475 $\cdot 10^{-30} $& 2.2884 $\cdot 10^{-17}$  & 2.2619 $\cdot 10^{-13}$ & 
1.5439 $\cdot 10^{-15}$ & 3.3240 $\cdot 10^{-16}$\hspace{1mm} \\ 
\hspace{1mm}& & & &  &\hspace{1mm}  \\
\hspace{1mm} $ n=15 $ &4.8677 $\cdot 10^{-35} $&8.9382 $\cdot 10^{-20}$  &3.4304 $\cdot 10^{-15}$  & 
1.0568 $\cdot 10^{-17}$ &1.7758 $\cdot 10^{-18}$ \hspace{1mm} \\
 \hspace{1mm}& & & &  &\hspace{1mm}  \\
\hspace{1mm} $ n = 16, $ &1.0336 $\cdot 10^{-39}$ &2.3871 $\cdot 10^{-22}$  &3.9063 $\cdot 10^{-17}$ & 
 6.9905 $\cdot 10^{-20}$ &9.4872 $\cdot 10^{-21}$ \hspace{1mm} \\ 
\hspace{1mm}& & & &  &\hspace{1mm}  \\
\hspace{1mm} $ n = 17 $ & 1.0984 $\cdot 10^{-44} $ & 4.3587 $\cdot 10^{-25}$ & 3.3395 $\cdot 10^{-19}$  & 
 2.4771 $\cdot 10^{-22}$ & 2.5443 $\cdot 10^{-23}$  \hspace{1mm} \\
 \hspace{1mm}& & & &  &\hspace{1mm}  \\
\hspace{1mm} $ n = 18  $ \,\, &\,\,  5.8409 $\cdot 10^{-50}$ \,\, &\,\,  5.4406 $\cdot 10^{-28}$ \,\, &\,\,  2.1431 $\cdot 10^{-21} $ \,\, &\,\,  
8.4825 $\cdot 10^{-25}$\,\,  &\,\,   6.8235 $\cdot 10^{-26}$  \hspace{1mm} \\ 
\hspace{1mm}& & & &  &\hspace{1mm}  \\
\hspace{1mm} $ n=19 $ &\,\, 1.5534 $\cdot 10^{-55}$ \,\, &\,\, 4.6420 $\cdot 10^{-31} $ \,\, &\,\, 1.0323 $\cdot 10^{-23}$  \,\, &\,\,  
1.5538 $\cdot 10^{-27}$\,\, &\,\,  9.1783 $\cdot 10^{-29}$   \hspace{1mm} \\
 \hspace{1mm}& & & &  &\hspace{1mm}  \\
\hspace{1mm} $ n=20$ &\,\,2.0604 $\cdot 10^{-61}$  \,\,&\,\,2.7070 $\cdot 10^{-34}$ \,\, &\,\, 3.7316 $\cdot 10^{-26}$ \,\, &\,\, 
2.7538 $\cdot 10^{-30} $\,\, &\,\, 1.2346 $\cdot 10^{-31} $\,\, \hspace{1mm} \\
\hspace{1mm}& & & &  &\hspace{1mm}  \\
\tableline   
\end{tabular}
\end{center}
\label{table:results1}
\end{table}

\newpage
\section{Appendix B}

In this appendix we present the expression for the EFP, which fits our 
exact data for $\alpha \geq \alpha_{c}$:
\be\label{EFP3} 
P_{>}(n) = \left[\sin\left(\frac{k_{F}^{-}- k_{F}^{+}}{2}\right)
\sin\left(\frac{k_{F}^{-}+ k_{F}^{+}}{2}\right)\right]^{\frac{n^{2}}{2}}
\times \left( \frac{A_{1}+(-1)^{n}B_{1}}{n^{1/2}}
+ \frac{C_{1} + D_{1} \cos(k_{F}^{+} n)} {n^{1/4}} \right).
\ee
For different values of the parameters $\alpha$ the fitting values of the coefficients $A_{1},...D_{1}$ are given in Table II. 
\vspace{10mm}
\begin{table}
\caption{The coefficients of the fitting formula (\ref{EFP3}) for the emptiness 
formation probability $P(n)$ for different values of the parameter $ \alpha > \alpha_{c}$.}
\begin{center}
\begin{tabular}{c|c|c|c|c|c|}
\tableline 
\hspace{1mm}& & & &  &\hspace{1mm}  \\
&$\alpha$  & $A_{1}$  &$B_{1} $ & $C_{1} $ & $ D_{1} $ \\
\hspace{1mm}& & & &  &\hspace{1mm}  \\
\tableline 
\hspace{1mm}& & & &  &\hspace{1mm}  \\
&\hspace{1mm}      2.005\hspace{1mm}  &\hspace{1mm}     0.155 \hspace{1mm} &  \hspace{1mm}   0.009 \hspace{1mm}& 
\hspace{1mm}    0.302\hspace{1mm} & \hspace{1mm}      0.314\hspace{1mm}  \\
&\hspace{1mm}      2.01 \hspace{1mm}  &\hspace{1mm}     0.204 \hspace{1mm}   & \hspace{1mm}  0.011 \hspace{1mm}& 
\hspace{1mm}    0.370 \hspace{1mm} & \hspace{1mm}     0.194\hspace{1mm} \\
&\hspace{1mm}      2.05  \hspace{1mm} &\hspace{1mm}     0.331 \hspace{1mm}   & \hspace{1mm}  0.018  \hspace{1mm}& 
\hspace{1mm}    0.350 \hspace{1mm}  &  \hspace{1mm}     0.072\hspace{1mm} \\
&\hspace{1mm}      2.1   \hspace{1mm} & \hspace{1mm}    0.351  \hspace{1mm}  & \hspace{1mm}  0.021  \hspace{1mm}& 
\hspace{1mm}    0.322 \hspace{1mm}  & \hspace{1mm}      0.063\hspace{1mm} \\
&\hspace{1mm}      2.2   \hspace{1mm} &\hspace{1mm}     0.572  \hspace{1mm}  & \hspace{1mm}  0.032 \hspace{1mm} & 
\hspace{1mm}     0.150 \hspace{1mm}   &   \hspace{1mm}   0.021\hspace{1mm} \\
&\hspace{1mm}      2.3   \hspace{1mm} &\hspace{1mm}     0.673 \hspace{1mm}   &\hspace{1mm}   0.037  \hspace{1mm}&  
\hspace{1mm}   0.064  \hspace{1mm}  &\hspace{1mm}     -0.003\hspace{1mm} \\
&\hspace{1mm}      2.4   \hspace{1mm} & \hspace{1mm}    0.683  \hspace{1mm}  & \hspace{1mm}  0.039  \hspace{1mm}&
\hspace{1mm}   0.045  \hspace{1mm}  &\hspace{1mm}     -0.006\hspace{1mm} \\
&\hspace{1mm}      2.5   \hspace{1mm} & \hspace{1mm}    0.685 \hspace{1mm}   & \hspace{1mm}  0.041  \hspace{1mm}&   
\hspace{1mm}  0.036   \hspace{1mm}  &\hspace{1mm}    -0.006\hspace{1mm} \\
&\hspace{1mm}       3    \hspace{1mm} &\hspace{1mm}     0.682  \hspace{1mm}  & \hspace{1mm}  0.045  \hspace{1mm}&   
\hspace{1mm}  0.012   \hspace{1mm}  &\hspace{1mm}    -0.005\hspace{1mm} \\
&\hspace{1mm}       4    \hspace{1mm} &\hspace{1mm}     0.675  \hspace{1mm}  &  \hspace{1mm} 0.050  \hspace{1mm}&  
\hspace{1mm}  -0.0002 \hspace{1mm}  &\hspace{1mm}     -0.004\hspace{1mm} \\
& \hspace{1mm}      10   \hspace{1mm} &\hspace{1mm}     0.662  \hspace{1mm}  & \hspace{1mm}  0.053  \hspace{1mm}&  
\hspace{1mm}  -0.006  \hspace{1mm}  &\hspace{1mm}     -0.003\hspace{1mm} \\
& \hspace{1mm}   $\infty$ \hspace{1mm} &  \hspace{1mm}     0.661  \hspace{1mm}  & \hspace{1mm}    0.054 \hspace{1mm} 
 & \hspace{1mm}   0 \hspace{1mm}   &  \hspace{1mm}   0 \hspace{1mm} \\
\hspace{1mm}& & & &  &\hspace{1mm}  \\
\tableline   
\end{tabular}
\end{center}
\label{table:results2}
\end{table}
\noindent

\end{document}